\documentclass[sigconf]{acmart}
\usepackage{graphicx}
\usepackage{multirow}
\usepackage{subfigure}
\usepackage{threeparttable}
\AtBeginDocument{%
  }

\acmYear{2025}\copyrightyear{2025}
\setcopyright{acmlicensed}
\acmConference[FSE '25]{33rd ACM International Conference on the Foundations of Software Engineering}{June 23--28, 2025}{Trondheim, Norway}
\acmBooktitle{33rd ACM International Conference on the Foundations of Software Engineering (FSE '25), June 23--28, 2025, Trondheim, Norway}
\acmDOI{10.1145/3696630.3728558}
\acmISBN{979-8-4007-1276-0/25/06}

\begin{document}

% \title{\name{}: Escalating and Assigning Tickets of Interest Using LLM}
\title{\name{}: Leveraging Large Language Models for Automated Ticket Escalation}

\author{Fengrui Liu}
\email{liufengrui.work@bytedance.com}
% \orcid{0000-0001-7578-0492}
\affiliation{%
  \institution{ByteDance}
  \city{Beijing}
  \country{China}
}

\author{Xiao He}
\email{xiao.hx@bytedance.com}
\affiliation{%
  \institution{ByteDance}
  \city{Hangzhou}
  \country{China}
}

\author{Tieying Zhang}
\email{tieying.zhang@bytedance.com}
\affiliation{%
  \institution{ByteDance}
  \city{San Jose}
  \country{United States}
}
\authornote{Corresponding author.}

\author{Jianjun Chen}
\email{jianjun.chen@bytedance.com}
\affiliation{%
  \institution{ByteDance}
  \city{San Jose}
  \country{United States}
}

\author{Yi Li}
\email{liyi.ly@bytedance.com}
\affiliation{%
  \institution{ByteDance}
  \city{Beijing}
  \country{China}
}

\author{Lihua Yi}
\email{yilihua@bytedance.com}
\affiliation{%
  \institution{ByteDance}
  \city{Shanghai} 
  \country{China}
}

\author{Haipeng Zhang}
\email{zhanghaipeng.zhp@bytedance.com}
\affiliation{%
  \institution{ByteDance}
  \city{Shanghai} 
  \country{China}
}

\author{Gang Wu}
\email{wugang.chewu@bytedance.com}
\affiliation{%
  \institution{ByteDance}
  \city{Shanghai}
  \country{China}
}

\author{Rui Shi}
\email{shirui@bytedance.com}
\affiliation{%
  \institution{ByteDance}
  \city{Beijing}
  \country{China}
}

\renewcommand{\shortauthors}{Fengrui Liu et al.}

\newcommand{\name}{\textit{TickIt}}

\begin{abstract}
In large-scale cloud service systems, support tickets serve as a critical mechanism for resolving customer issues and maintaining service quality. 
However, traditional manual ticket escalation processes encounter significant challenges, including inefficiency, inaccuracy, and difficulty in handling the high volume and complexity of tickets. 
While previous research has proposed various machine learning models for ticket classification, these approaches often overlook the practical demands of real-world escalations, such as dynamic ticket updates, topic-specific routing, and the analysis of ticket relationships. 
To bridge this gap, this paper introduces \name{}, an innovative online ticket escalation framework powered by Large Language Models. 
\name{} enables topic-aware, dynamic, and relationship-driven ticket escalations by continuously updating ticket states, assigning tickets to the most appropriate support teams, exploring ticket correlations, and leveraging category-guided supervised fine-tuning to continuously improve its performance. 
By deploying \name{} in ByteDance's cloud service platform Volcano Engine, we validate its efficacy and practicality, marking a significant advancement in the field of automated ticket escalation for large-scale cloud service systems.
\end{abstract}





\maketitle

\section{Introduction}
With the rapid development of cloud computing technologies, an increasing number of applications are either being migrated to the cloud or built to run natively on the cloud from day one.
For cloud service vendors, support tickets serve as an important method of facilitating communication between customers and support analysts.
When customers submit support tickets, they typically express the issues using natural language, encompassing a range of inquiries such as usage questions, feature requests, bug reports and system failures.
Support analysts then respond to the tickets or initiate chat sessions with customers to resolve these issues.
Generally, most tickets can be effectively closed upon resolution of the reported issues.
However, when critical issues arise, such as severe system incidents or intense customer complaints, it is essential to promptly escalate the tickets to on-call engineers or customer managers.
With the volume of thousands of tickets each day, manual escalations heavily rely on the experience of support analysts, which can result in erroneous escalations or delays\cite{ling_predicting_2005}.
Thus, an automated online ticket escalation method is essential for enhancing the efficiency and accuracy of the customer support teams.
In this study, we analyze over 20,000 tickets from Volcano Engine\cite{noauthor_volcano_nodate}, the public cloud platform of ByteDance, and share observations and our practical experiences of deploying a ticket escalating system as follows.

First, the issues addressed in support tickets can vary significantly, requiring prompt and accurate escalation to the appropriate support teams based on their topics. 
For on-call engineers, they prioritize critical system incidents to reduce Mean Time to Repair (MTTR) and improve Service Level Agreements (SLAs).
While customer managers focus on addressing intense customer complaints to enhance satisfaction and retention.
As for security engineers who are tasked with safeguarding the platform against potential threats, concentrate specifically on security incidents.
Therefore, understanding the topics of tickets and appropriately escalating them to relevant supporters is necessary.
However, existing binary classification models\cite{montgomery_customer_2020,gupta_hybrid_2020,pavelski_real-world_2022}, which determine whether to escalate, fail to route tickets to the appropriate teams based on their content.
Based on our observations, it is vital to predefine distinct ticket categories according to the responsibilities and interests of the support teams. 
Moreover, the state of a ticket changes constantly throughout the dialogue between customers and support analysts.
Rather than being completely provided when the ticket is created, some important details are clarified during their conversations, emphasizing the need for an online ticket escalation method.
Some existing methods\cite{montgomery_customer_2020,molino_cota_2018,marcuzzo_multi-level_2022} only classify tickets once, overlooking potential changes in ticket states that might necessitate an escalation.
We observed that continuously analyzing the latest conversations between customers and support analysts with an online manner can help keep ticket states updated, facilitating timely and accurate ticket escalations.
This allows us to perform multi-classification of tickets to facilitate the online escalations.

% 升级如何关联多个oncall内容
Second, when series issues arise, such as multiple customers encounter the same system failure, similar tickets are often submitted by different customers.
Analyzing common topics across various tickets can help support analysts in understanding the severity of the issues.
By examining the topics among similar tickets, support analysts can further assess the impact of the issues and make informed decisions regarding escalations.
Compared to the existing methods\cite{feng_tadaa_2023,liu_ticket-bert_2023} that analyze tickets individually, we observe that uncovering the relationships among tickets can provide a comprehensive overview, ensuring the overall stability of the cloud platform and mitigating the risk of overlooking critical issues due to content biases in individual tickets.
Moreover, by consolidating escalations of similar tickets, we can further enhance the efficiency of support analysts and reduce operational costs.

Third, the analysis of natural language-based tickets presents significant challenges.
Existing ticket escalation methods that rely on feature engineering\cite{werner_how_2018, werner_can_2019} often struggle to effectively comprehend the content of tickets, thereby leading to difficulties in accurately identifying critical issues and escalating tickets in real-world applications. 
Although large language models (LLMs)~\cite{brown_language_2020} have recently gained popularity due to their remarkable performance in natural language processing across various domains, research on leveraging LLMs specifically for ticket escalation remains limited.
Benefiting from their strong generalization capabilities, LLMs can be adapted to specific target tasks.
However, a key challenge lies in effectively utilizing feedback to continuously optimize the performance of LLM-based methods.
Further investigations are needed to enhance their capabilities through advanced techniques, including non-parametric prompt engineering\cite{wei_chain--thought_2022,brown_language_2020} and parameter-efficient fine-tuning methods\cite{hu_lora_2021}.
These approaches hold the potential to significantly enhance the effectiveness of LLMs in ticket escalation tasks. 
By improving the understanding of ticket content, such advancements could lead to the development of more robust and efficient ticket escalation strategies.

To tackle the above challenges, we propose \name{}, a framework for escalating customer tickets in Volcano Engine.
In particular, \name{} predefines various categories of tickets based on the responsibilities and interests of support analysts.
It then continuously follows the latest conversations between customers and support analysts with an online manner to maintain ticket states updated.
By utilizing large language models\cite{zhao_survey_2024}, \name{} comprehends the ticket topics and mines the connections among tickets to facilitate the escalations.
The salient contributions of our work are as follows:

% \begin{itemize}
%   \item We propose a novel online ticket escalation method \name{} that leverages a large language model to comprehend the latest ticket content, enabling on-demand escalation to support analysts who are responsible or interested in them.
%   \item We further utilize finite state machines to model the state of tickets and explore connections among them, which helps prevent missed or redundant escalations.
%   \item Based on user feedback regarding escalations, we enhance offline supervised fine-tuning (SFT) with category guidance. It can specifically improve the accuracy of ticket escalations.
%   \item We conduct extensive experiments to evaluate the proposed method \name{}, which achieves an $F1$ score of 0.862 in an online production environment. Additionally, we report ablation studies to validate the effectiveness of the \name{} design.
% \end{itemize}

\begin{itemize}
  \item We systematically reveal the nature of practical ticket escalation in cloud service systems, emphasizing the necessity of online, topic-aware, and relationship-driven escalations to improve efficiency and accuracy in customer support.
  \item We propose \name{}, an end-to-end ticket escalation framework based on large language models. \name{} continuously updates ticket states, explores relationships among tickets, and integrates category-guided fine-tuning based on user feedback, significantly improving escalation accuracy and efficiency.
  \item We deploy \name{} in the production environment of Volcano Engine's ticket management system and demonstrate its effectiveness through extensive experiments and real-world cases, achieving significant improvements in ticket escalation accuracy and efficiency.
\end{itemize}

The remainder of this paper is organized as follows. Section \ref{sec:related_work} reviews the related work. Section \ref{sec:methodology} presents the methodology of our proposed method \name{}. Section \ref{sec:experiments} reports the experiments and results. Section \ref{sec:discussion} discusses the results and implications, and Section \ref{sec:conclusion} concludes the paper and outlines future work.

\begin{figure*}[t]
    \centering
    \includegraphics[width=\textwidth]{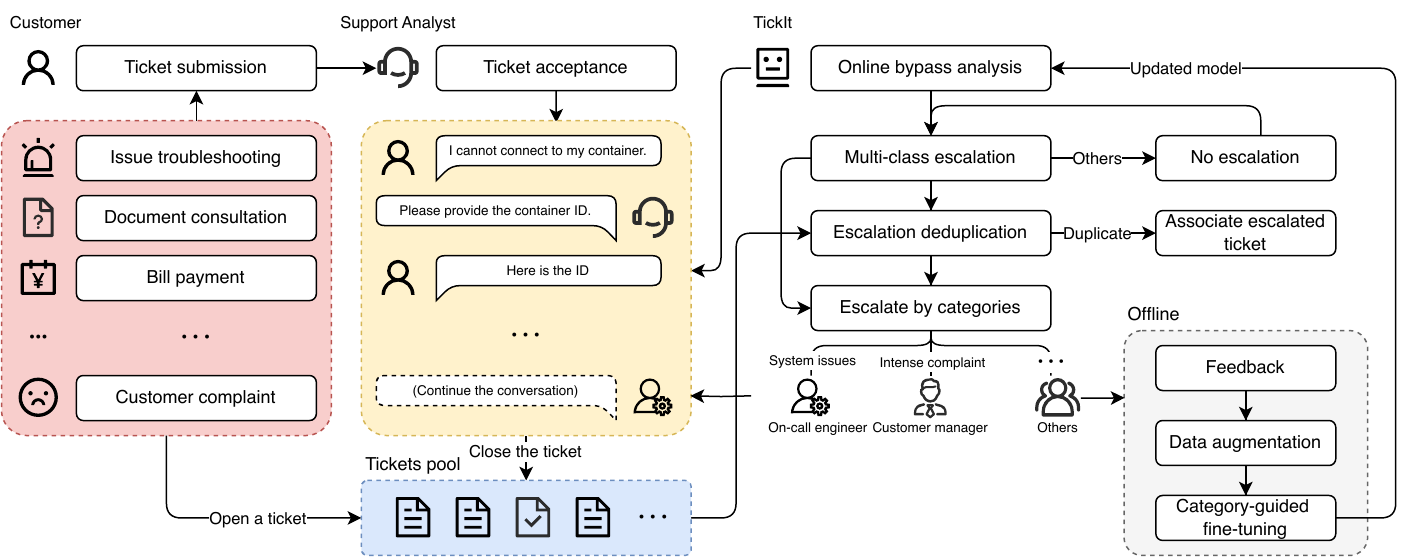}
    \caption{Overview framework of \name{}.}
    \label{fig:overview}
\end{figure*}

\section{Related Work}
\label{sec:related_work}

Escalating and addressing severe customer tickets promptly is highly advantageous for Volcano Engine, as it directly impacts service quality and customer satisfaction\cite{yang_predictive_2022}.
This section begins with a review of relevant work on the application of ticket escalation, followed by an introduction to the application of language models in this domain.

\textbf{Automatic Customer Tickets Escalation.}
To mitigate the risks associated with missed escalations and delays in manual ticket handling\cite{ling_predicting_2005}, researchers have proposed various automated methods for ticket escalation\cite{montgomery_customer_2020,marcuzzo_multi-level_2022,gupta_hybrid_2020,pavelski_real-world_2022,chen_towards_2020}.
Some studies formalize this task as a binary classification problem\cite{werner_how_2018, pavelski_real-world_2022}, where the objective is to determine whether a ticket should be escalated or not.
However, in our real-world cloud platform, Volcano Engine, different analysts in distinct roles are interested in different types of ticket topics.
These binary classification methods overlook the specific topics of tickets, rendering it impossible to escalate tickets to analysts accordingly.
Recognizing this limitation, we advocate for a multi-class classification approach\cite{molino_cota_2018,marcuzzo_multi-level_2022}, which aligns more closely with the needs of cloud platforms.

Regarding the models used for automatic ticket escalation, some methods involve extracting features such as enumerated values, word frequency\cite{molino_cota_2018} from ticket titles, descriptions \cite{gupta_hybrid_2020,janusz_predicting_2020} and customer profiles \cite{montgomery_customer_2020}.
These features are then utilized to train classification models such as extreme gradient boosting tree\cite{gupta_hybrid_2020,janusz_predicting_2020}, random forests\cite{montgomery_what_2017}, and support vector machines\cite{montgomery_customer_2020}.
However, such feature engineering methods\cite{werner_how_2018, werner_can_2019} heavily rely on the selected features for the classification model and exhibit limitations in semantic understanding of ticket content. 
Recent research has highlighted the necessity of enhancing content comprehension in ticket analysis\cite{lin_opinion_2022}.
Some studies successfully employ word embeddings\cite{marcuzzo_multi-level_2022} and language models\cite{liu_ticket-bert_2023,pavelski_real-world_2022,feng_tadaa_2023} to understand customer intent effectively, thereby avoiding the manual selection of ticket features and making full use of the textual information contained in customer tickets.

To minimize the costs on tickets escalation, some studies\cite{zheng_ifeedback_2019,liu_incident-aware_2023} aggregate similar tickets according to the topics\cite{ponay_topic_2022} from a large number of customer tickets and reduce redundant escalations.
iFeedback\cite{zheng_ifeedback_2019} propose an aggregation method using feature vectors while exploring user feedback.
iPACK\cite{liu_incident-aware_2023} correlates customer tickets with issues relevant to the cloud platform.
Although these studies are primarily concerned with the analysis of cloud platform failures, they extend beyond individual tickets to explore the interrelationships among multiple tickets.
By aggregating duplicate tickets, the overall number of escalations can be effectively reduced.

\textbf{Language Models for Tickets Analysis.}
Customer tickets are typically expressed in natural language, encompassing elements such as ticket titles, detailed descriptions, and dialogues between customers and support analysts.
This presents a challenge in understanding the underlying topics of these tickets \cite{truss_ai-based_2024}.
Recent advancements in natural language processing prompt researchers to leverage language models for customer tickets analysis.
The proposal of bidirectional encoder representations from transformers (BERT)\cite{devlin_bert_2019} has significantly transformed this landscape.
Supp-BERT\cite{marcuzzo_multi-level_2022} enhances word embeddings by utilizing contextualized representations from BERT, thereby improving the extraction and classification of ticket content.
BERTopic\cite{ponay_topic_2022,grootendorst_bertopic_2022} employs topic modeling techniques to systematically categorize tickets, showcasing the effectiveness of language models.
Ticket-BERT\cite{liu_ticket-bert_2023} further fine-tunes a pre-trained BERT model on historical customer tickets, enhancing the accuracy for customer tickets classification.
Benefiting from the contextual understanding of textual tickets, these BERT-based ticket analysis approaches\cite{marcuzzo_multi-level_2022,grootendorst_bertopic_2022,liu_ticket-bert_2023} significantly outperform the feature engineering-based methods\cite{werner_how_2018, werner_can_2019}.

In recent years, with the widespread popularity of generative pre-trained transformer (GPT)\cite{vaswani_attention_2017}, autoregressive large language models demonstrate remarkable capabilities in contextual comprehension\cite{brown_language_2020}, leading to significant advancements in general natural language processing\cite{zhou_comprehensive_2023}.
However, only a limited number of researchers\cite{arici_llm-based_2023} explore GPT for specific applications in ticket escalation.
In this paper, we contribute to this area by studying the application of GPT within the context of ticket escalations.
We validate the effectiveness of various methodologies aimed at enhancing performance, including advanced techniques such as Chain of Thought (CoT) prompting\cite{wei_chain--thought_2022}, in-context learning (ICL)\cite{brown_language_2020,min_rethinking_2022,dai_why_2023} and supervised fine-tuning (SFT) approaches\cite{hu_lora_2021}.
Our findings aim to bridge the gap in current research and provide insights into the potential of large language models for improving the efficiency and accuracy of customer tickets escalations.

\section{Methodology}
\label{sec:methodology}

In this section, we introduce \name{}, a framework to enhance the effectiveness of tickets escalation in Volcano Engine.
As illustrated in Figure \ref{fig:overview}, when customers encounter issues while using the cloud platform, they submit a support ticket, which initiates a chat session with support analysts.
At this stage, \name{} operates in a bypass mode to access the latest conversation content in real-time and perform several key functions:
(1) \textbf{Multi-class ticket escalation:} Utilizing a large language model as a classifier to assess the necessity of escalating the ticket.
(2) \textbf{Escalation deduplication:} Identifying and escalating issues that have not previously been addressed.
(3) \textbf{Model fine-tuning:} Augmenting ticket data and fine-tuning the model to enhance its performance based on feedback.

\subsection{Multi-class ticket escalation}
\label{subsec:multi_class_ticket_escalation}

In Volcano Engine, customers submit tickets for various reasons when they encounter issues, such as system failures, document consultation, bill payment, or customer complaints.
During the dialogues between customers and customer analysts, the details of these issues can be further clarified.
When a serious issue is identified, it is crucial to escalate the customer ticket appropriately.
Different support analysts have distinct responsibilities and priorities regarding ticket topics.
For instance, on-call engineers focus on critical system failures, which are vital for improving Service Level Agreements (SLAs).
While customer managers prioritize intense complaints to enhance satisfaction and retention.
Therefore, escalated customer tickets should be assigned to the appropriate analysts responsible for addressing them.

In \name{}, we formalize the aforementioned problem as a multi-class classification task concerning the content of customer tickets.
Based on different responsibilities of support analysts, we predefine several categories of ticket topics, such as system failure, customer complaint, and asset loss.
These ticket categories can be flexibly configured according to the responsibilities and interests of support analysts.
Specifically, we also apply an exclusion method to categorize any tickets that do not belong to the predefined categories as \textit{"Others"}, which do not require escalation.

\begin{figure}[t]
    \centering
    \includegraphics[width=1\linewidth]{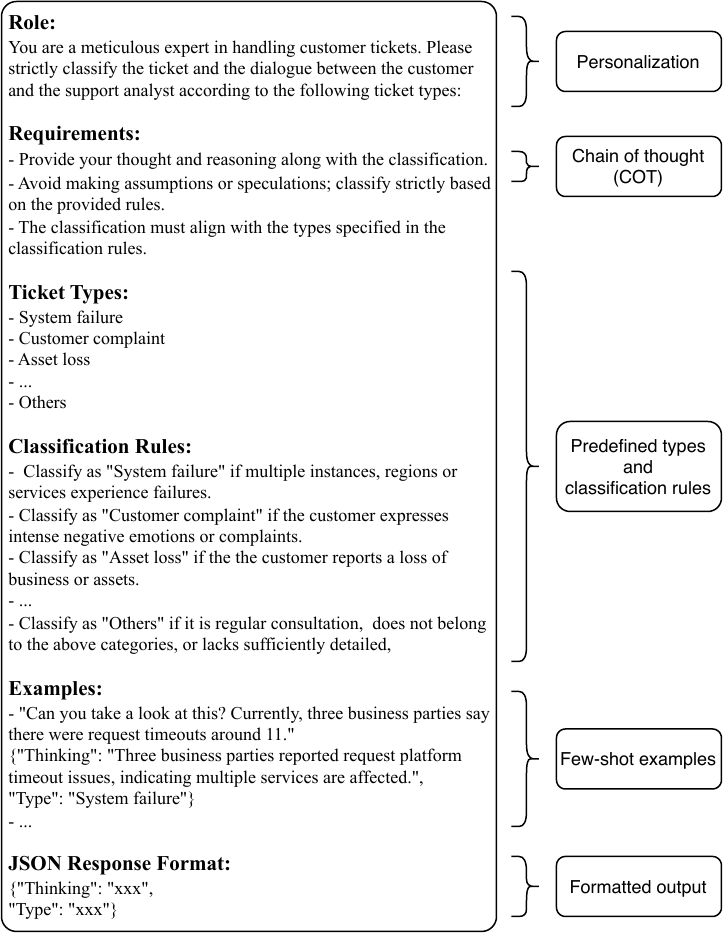}
    \caption{Classification prompt for customer tickets escalation}
    \label{fig:escalation_prompt}
\end{figure}

To enhance the performance of the large language model in ticket escalation, the system prompt, as shown in Figure \ref{fig:escalation_prompt}, includes the following key components.
First, it defines the role\cite{wang_rolellm_2024} of the large language model, allowing it to undertake the given classification task while conforming to personalized role characteristics\cite{tao_rolecraft-glm_2024,wang_rolellm_2024}.
Role-playing is regarded as a method that allows large language models to demonstrate specific personality traits, which can effectively enhance the quality of their responses. 
In \name{}, we designate the model as a meticulous expert of our cloud platform, making it closely aligned with predefined ticket types for effective classification.
Next, we employ the Chain of Thought (CoT) technique\cite{wei_chain--thought_2022} to further improve the accuracy of the large language model in ticket classification.
This technique involves explicitly prompting the large language model to outline its reasoning before answering the final classification result. 
By encouraging the model to think through its decision-making process, it maintains logical consistency and promotes a comprehensive understanding of the ticket content.
This structured reasoning not only organizes the responses effectively, but also enhances the trustworthiness of the classification results. 
Support analysts are more likely to trust these results when they can follow the reasoning steps, rendering the outcomes explainable and reliable.

Subsequently, we provide few-shot examples to assist the LLM in comprehending the ticket categories, a process known as In-Context Learning (ICL)\cite{brown_language_2020}.
ICL represents a novel paradigm of analogical learning for prompt engineering, it enables models to learn and reason through a limited number of labeled samples, and enhance their learning performance.
To further automate the parsing of outputs from the large language models, we instruct the model to generate results in a structured format\cite{guzman_introduction_2024} within the system prompt.
These structured output with specific schema enhance the robustness for downstream tasks that rely on these results.

\subsection{Escalation deduplication}
\label{subsec:escalation_deduplication}

In Section \ref{subsec:multi_class_ticket_escalation}, we introduce the methodology employed by a large language model to assess whether an individual customer ticket warrants escalation.
Once a ticket is selected to be escalated, \name{} needs to review all currently opened tickets within the tickets pool to check if any similar issues have previously been escalated.
For example, a system fault may affect multiple customers, resulting in several customer tickets being submitted. 
By identifying tickets that related to the same issue, we can significantly minimize redundant escalations without overlooking the system fault, thereby enhancing the operational efficiency of support analysts.

\begin{figure}[t]
    \centering
    \includegraphics[width=1\linewidth]{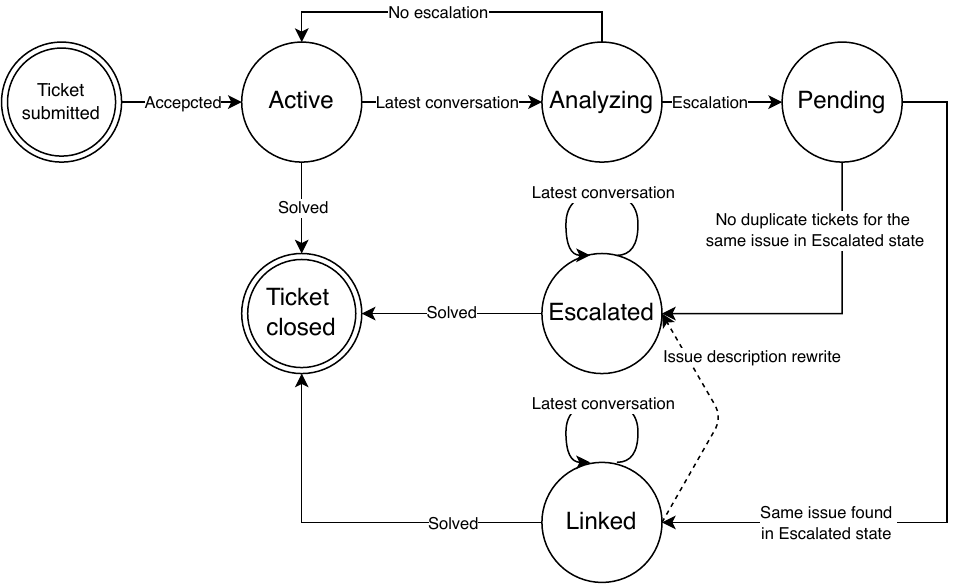}
    \caption{Customer ticket state within its lifecycle}
    \label{fig:ticket_state}
\end{figure}

To achieve the above goals, we formally represent the states of a customer ticket within its lifecycle as a finite state machine, as illustrated in Figure \ref{fig:ticket_state}.
Once the customer ticket is accepted, it transitions into an active state.
Whenever a customer engages in a conversation with support analyst, the content of the latest dialogue triggers \name{} start a new round of analysis, transitioning the ticket into the analyzing state.
At this point, \name{} utilizes the method described in Section \ref{subsec:multi_class_ticket_escalation} to determine whether the current ticket requires escalation.
If the current ticket is classified as \textit{"Others"}, indicating that escalation is unnecessary, its state reverts to active, awaiting the next round of interaction.
Conversely, if the ticket is classified as a predefined ticket type, it enters the pending state to check whether a similar ticket has been escalated previously.

\begin{figure}
    \centering
    \subfigure[Prompt for identifying ticket issues.]{
        \includegraphics[width=1\linewidth]{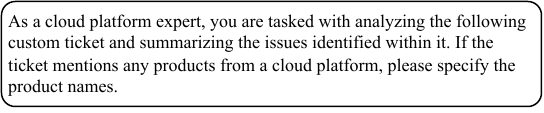}
        \label{subfig:summary_prompt}
    }
    \subfigure[Prompt for rewriting escalated ticket issues.]{
        \includegraphics[width=1\linewidth]{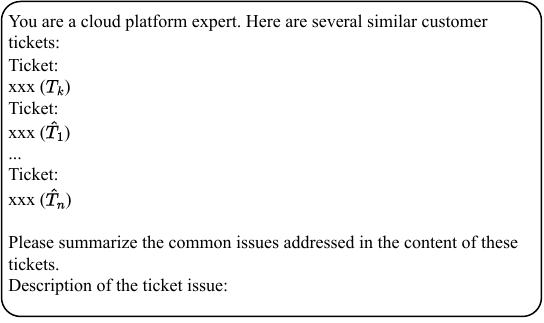}
        \label{subfig:rewriting_prompt}
    }
    \caption{Prompt for escalation deduplication.}
    \label{fig:deduplication}
\end{figure}

For deduplicating tickets, it is essential to identify the specific issues described within them, rather than focusing on other ticket information.
Based on this observation, we propose a deduplication method that leverages the extracted issues from the tickets.
When a ticket is in the pending status, we use the prompt shown in Figure \ref{subfig:summary_prompt} to instruct the large language model to summarize the issue descriptions mentioned in the ticket, and to identify the corresponding cloud products as accurately as possible.
Next, \name{} retrieves all tickets that are currently in escalated state from the ticket pool.
Since these escalated tickets have transitioned from the pending state, they also contain issue descriptions of the tickets.
\name{} needs to analyze the similarities between the current ticket issue and those of escalated tickets in order to avoid duplicate escalations.
Although the issues of escalated tickets are expected to be unique, it is still inevitable to have a significant number of tickets in a large-scale cloud service platform.
To address this, \name{} first uses an embedding model to convert the ticket issue \(T\) into an embedding vector \(v\) as
\begin{equation}
v = f_{Embedding}(T)
\end{equation}
\noindent It then compares the current embedding vector \(v\) for similarity with the embedding vectors \(\{v_1,v_2,...,v_n\}\) of all tickets \(\{T_1,T_2,...,T_n\}\) that are in escalated state, selecting the vector \(v_k\) that exhibits the highest similarity as
\begin{equation}
v_k = arg max_{v_i\in \{v_1,v_2,...,v_n\}}\frac{v\cdot v_i}{\left \|  v\right \| \left \| v_i \right \| }
\end{equation}
\noindent If the most similar result \(\frac{v\cdot v_k}{\left \|  v\right \| \left \| v_k \right \| }\) exceeds a specified threshold \(\theta\), the current ticket can be marked as escalated state.
All escalated tickets are transferred through the above process, thus the issues they represent are unique.
Otherwise, it is considered to be linked with an existing similar escalated ticket.
Note that if an escalated ticket is linked by other tickets, it represents a specific class of issues.
To enhance its representational capability for this issue class and eliminate bias in issue descriptions, we employ large language model in Figure \ref{subfig:rewriting_prompt} to rewrite the issue description of the escalated ticket,
\begin{equation}
\tilde{T_k} = f_{Rewriting}(T_k, \hat{T}_1,\hat{T}_2,...\hat{T}_n)
\end{equation}
\noindent where \(T_k\) denotes the selected ticket in escalated state, while \(\hat{T}_n\) represents all tickets linked to \(T_k\). 
The new description \(\tilde{T}\) is used to encapsulate the rewritten description \(T\) of the escalated ticket.
This process aims to highlight the commonalities among these tickets.

The lifecycle of all tickets ends when the customer closes them, which typically indicates that the issues have been resolved.
At this point, \name{} removes the closed tickets from the ticket pool and no longer considers them in the tickets deduplication process. 
This practice helps maintain a manageable ticket pool size and allows us to identify recurring issues over time.

\subsection{Category-guided fine-tuning}
\label{sec:finetune}

When a ticket is escalated as Section \ref{subsec:escalation_deduplication} described, \name{} sends an alert notification to the corresponding analyst according to the ticket type.
This notification contains a summary of the ticket issue that is generated, as Figure \ref{subfig:summary_prompt} shows, and features three interactive buttons. 
Two of these buttons allow the analyst to upvote or downvote the current alert, facilitating the assessment of its validity. 
The third button provides a link to the ticket, enabling the analyst to be redirected to the associated chat group.
By recording the interactions of the analysts with these notifications, \name{} utilizes these records as feedback for the automatic escalations.
Specifically, we consider upvotes and downvotes as the highest priority feedback, and treat them as direct labels for ticket escalations.
In cases where the support analyst does not provide either type of feedback, we check whether they join the ticket group chat session via the redirect button.
If the notified analyst joins the ticket handling process via the redirect link, we consider it as an indirect label indicating that the escalation is deemed appropriate.

After collecting extensive feedback based on this method, \name{} aims to utilize these data to improve the multi-class classification of tickets during the escalation process.
While there are existing researches\cite{schulman_proximal_2017,ouyang_training_2022,rafailov_direct_nodate} utilize Reinforcement Learning from Human Feedback (RLHF) to enhance the quality of outputs generated by large language models, such methods present limitations in the context of \name{}.
Take Direct Preference Optimization (DPO)\cite{ouyang_training_2022} method as an example, it relies on labeling multiple outputs from large language models to create preference pairs.
However, in \name{}, escalated tickets are only labeled once, which lacks preferences across different generations.
Consequently, \name{} is more suited for Supervised Fine-Tuning (SFT) method than RLHF for fine-tuning to enhance its performance on ticket escalations.

\begin{figure}
    \centering
    \includegraphics[width=\linewidth]{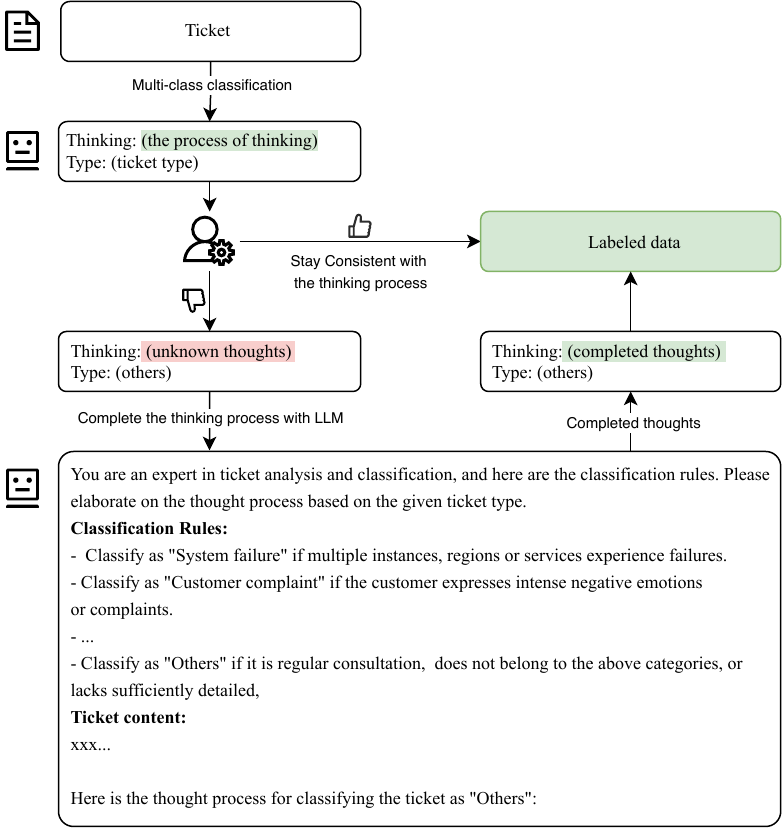}
    \caption{Data augmentation for labeled data.}
    \label{fig:data_augmentation}
\end{figure}

For positive upvote feedback, we directly utilize the outputs generated by the large language model as labels for the corresponding tickets.
Conversely, negative downvote feedback is interpreted as incorrect escalations of the tickets.
However, as discussed in Section \ref{subsec:multi_class_ticket_escalation}, human feedback can only correct the category of the tickets, as it lacks the thinking steps required by the Chain of Thought (CoT) technique. 
To address this limitation, we adopt the ticket types derived from customer feedback as the ground truth. 
We then employ the prompt shown in Figure \ref{fig:data_augmentation} to guide the large language model in completing the classification thinking steps for classifying the given ticket types.
Due to the diversity of thinking steps, we further conduct three sampling iterations of these possible thinking steps for each ticket to enrich the dataset.
Subsequently, we effectively process user feedback and apply a category-guided approach to data augmentation, ultimately constructing a comprehensive labeled dataset.
After accumulating a substantial amount of labeled data, we perform offline optimization of the model using the SFT method and subsequently update the online model to enhance classification performance on ticket escalations.

\section{Experiments}
\label{sec:experiments}

\subsection{Deployment and Dataset Collection}

The \name{} system has been deployed in production for the cloud service provider Volcano Engine since August 1st, 2024. 
By December 31st, 2024, \name{} had processed a total of 20,066 tickets and 654,901 messages. Among these, 2,012 tickets were escalated by \name{} and subsequently forwarded to the relevant analysts groups. 
Feedback, which includes both upvotes and downvotes, was received for these tickets, with approximately 81\% of the feedback indicating that the escalations were appropriate.

Prior to deployment, a dataset comprising 148 tickets was collected and manually labeled offline with the assistance of support analysts. 
This dataset, referred to as "Training Data I", is utilized to validate the proof of concept through prompt tuning during the initial deployment phase.
Notably, approximately $70\%$ of the tickets are labeled as "should not escalate." 
This dataset serves as the foundation for training.
Additionally, online escalated tickets with feedback are divided into two subsets for evaluation: tickets submitted between August 1st, 2024 and October 31st, 2024, are used as "Training Data II", while the remaining tickets are reserved for testing. 
Table \ref{tab:ticket_data} provides a detailed summary of the collected data and their respective splits for evaluation.

We select Doubao\cite{volcano_engine_doubao-volcano_nodate}, a large language model developed by ByteDance and hosted on Volcano Engine, as the foundational model of \name{}.
Initially, \name{} was deployed with a Chain-of-Thought (CoT) prompt, as shown in Figure \ref{fig:escalation_prompt}, which is tuned using "Training Data I." 
Following the collection of both parts of the training data, we apply the data augmentation technique introduced in Section \ref{sec:finetune} to enhance the datasets by incorporating the correct "Thoughts" for each ticket along with their corresponding types.
These augmented datasets are then used to perform supervised fine-tuning with LoRA~\cite{hu_lora_2021} on the base LLM model. 
Subsequently, an upgraded version of \name{} which utilizes the same prompt was deployed starting on November 1st, 2024. 
During this phase, both versions were simultaneously used online. 
A ticket was escalated if either version determined it should be.

\begin{table}[t]
\centering
\caption{Summary of ticket data for \name{} evaluation.}
\label{tab:ticket_data}
\begin{tabular}{@{}lr@{}}
\toprule
\textbf{Description}                               & \textbf{Count} \\ \midrule
\textbf{Offline Data (Prior to Deployment)}        &                \\
Training Data I (Offline tickets collected)                          & 148            \\ \midrule
\textbf{Total Data Processed (Aug 1st, 2024 – Dec 31st, 2024)} &                \\
Total tickets processed                            & 20,066         \\
Total messages processed                           & 654,901        \\ \midrule
\textbf{Escalated Tickets}                         &                \\
Escalated tickets                                  & 2,012          \\
Tickets with feedback (upvotes/downvotes)          & 472            \\ \midrule
\textbf{Evaluation Data Split}                     &                \\
Training Data II (Aug 1st, 2024 – Oct 31st, 2024)         & 312 \\
Testing Data \phantom{ II } (Nov 1st, 2024 – Dec 31st, 2024)          & 160 \\ \bottomrule
\end{tabular}
\end{table}

\subsection{Experimental Setups}
We evaluate two primary scenarios in the automated ticket escalation system: ticket escalation and escalation deduplication. While ticket escalation has been extensively studied in prior research, escalation deduplication remains relatively unexplored. Therefore, we conduct comprehensive experiments comparing our approach with various baselines for ticket escalation. For escalation deduplication, we perform ablation studies to systematically evaluate the performance and assess the contribution of individual components.

\subsubsection{Baselines}
We conduct evaluations using both small language model (SLM)-based methods and large language model (LLM)-based methods, including approaches based on prompt engineering as well as fine-tuning techniques. 

\textbf{SLM-based Methods.} A pre-trained small language model, such as BERT~\cite{liu_ticket-bert_2023}, is used to embed the messages contained within the tickets into dense feature vector representations. 
These representations are then fed into a traditional machine learning classifier, including logistic regression (LR), multi-layer perceptron (MLP) and gradient boosting decision trees (GBDT), which are trained on the labeled training data.

\textbf{LLM-based Methods.} We evaluate several widely adopted prompt engineering-based methods, including Chain of Thought (CoT)~\cite{wei_chain--thought_2022} and In-Context Learning (ICL)~\cite{brown_language_2020}, both of which are designed to enhance reasoning capabilities. 
For ICL, we adopt a Retrieve-Augmented Generation pipeline, where the most similar tickets to the current messages are retrieved using an embedding-based similarity model and subsequently incorporated into the prompt to facilitate in-context learning.
Additionally, we assess the Reflection method~\cite{shinn_reflexion_2023}, which focuses on self-correcting the reasoning and answers of LLM. 
Furthermore, we evaluate fine-tuning-based methods, specifically the proposed \name{} system, which leverages data augmentation and supervised fine-tuning, as introduced in Section \ref{sec:finetune}.

\begin{table*}[t]
\centering
\caption{Tickets escalation comparison across different methods.}
\begin{tabular}{c|l|ccc}
\toprule
\textbf{Category} & \textbf{Methods} & \textbf{Precision} & \textbf{Recall} & \textbf{F1-score} \\
\midrule
\multirow{3}{*}{\centering \shortstack{\textbf{SLM}}} 
 & Embed Vec. + LR & 0.788 & 0.744 & 0.765 \\
 & Embed Vec. + MLP & 0.753 & 0.856 & 0.801 \\
 & Embed Vec. + GBDT & 0.732 & 0.720 & 0.728 \\
\midrule
\multirow{4}{*}{\centering \shortstack{\textbf{LLM}\\\textbf{+}\\\textbf{Prompt}}} 
 & CoT & 0.817 & 0.821 & 0.819 \\
 & CoT + Reflection & \textit{0.828} & 0.824 & \textbf{0.826} \\
 & CoT + ICL & 0.810 & 0.892 & 0.849 \\
 & CoT + ICL + Reflection & 0.813 & 0.874 & 0.843 \\
\midrule
\multirow{4}{*}{\centering \shortstack{\textbf{LLM}\\\textbf{+}\\\textbf{Prompt}\\\textbf{+}\\\textbf{Fine-tune}}} 
 & SFT + CoT  & 0.818 & \textit{0.912} & \textit{\textbf{0.862}} \\
 & SFT + CoT + Reflection & 0.804 & 0.880 & 0.841 \\
 & SFT + CoT + ICL & 0.804 & 0.901 & 0.850 \\
 & SFT + CoT + ICL + Reflection & 0.804 & 0.907 & 0.852 \\
\bottomrule
\end{tabular}

\label{tab:slm_llm_result}
\end{table*}

\subsubsection{Evaluation metrics}
We detail the evaluation metrics of \name{} in ticket escalation and deduplication.

\textbf{Metric for evaluating ticket escalation}. 
In our approach, each ticket is represented as a sequence of messages, and we assign one of predefined ticket types to it. 
While these predefined categories provide flexibility to our method, they also raise difficulties for data labeling.
In our experiments, tickets classified as \textit{"Others"} are designated as non-escalated, while all other classifications trigger an escalation.
We leverage user feedback to establish the ground truth regarding the escalation status of tickets within the datasets under consideration.
Accordingly, we use \textit{Precision}, \textit{Recall}, and \textit{F1-score} as metrics to assess the performance of our proposed ticket escalation method.

\textbf{Metric for evaluating escalation deduplication}. 
In \name{}, each escalated ticket and its associated linked tickets are considered to represent the same type of issue.
Within a specified evaluation period, we regard each ticket that in escalated state as one group, while tickets that have not been escalated are grouped together as another category.
We utilize classification metrics \textit{Precision}, \textit{Recall}, and \textit{F1-score} to assess the effectiveness of \name{} in escalation deduplication.

\subsubsection{LLM setup}
We evaluate both open-sourced and commercial large language models. 
Since ticket messages contain private and sensitive information from customers of Volcano Engine, our evaluation of commercial models is restricted to the base LLMs from the Doubao family, specifically those hosted on Volcano Engine.
Specifically, we evaluate \texttt{Doubou-Pro-32k-20240615}, hereafter referred to as \texttt{Doubao-Pro}, unless otherwise specified.

In addition, we assess two open-sourced LLM models: Qwen-2.5 and LLaMA-3.1. 
For LLaMA-3.1, we use the instruct versions with 8 billion and 70 billion parameters, denoted as \texttt{Llama3.1-8B} and \texttt{Llama3.1-70B}, respectively. 
For Qwen-2.5, we use the instruct versions with 7 billion and 72 billion parameters, denoted as \texttt{Qwen2.5-7B} and \texttt{Qwen2.5-72B}, respectively.
This diverse selection of pre-trained models enables a comprehensive comparison of performance across different architectures and configurations.

\subsubsection{Implementation details}
All experiments involving \texttt{Doubao-Pro} are conducted on the Volcano Engine platform, encompassing both model evaluation and supervised fine-tuning. 
For LLM inference, we set the temperature to zero to ensure that the results are as reproducible as possible.
For supervised fine-tuning with LoRA, we adopt the settings with $lora\_alpha$ is 32, $lora\_rank$ is 32, and a learning rate of $5e\text{-}5$, while specifically configuring the training to run for 10 epochs with a batch size of 1.

For the remaining experiments involving open-sourced LLM models and SLM-based methods, we conduct experiments on a local machine equipped with 8 \textit{ NVIDIA A100 80GB GPUs}. The MLP used in the SLM baseline consists of three hidden layers with sizes 384, 128, and 64, respectively. 
The activation function employed is \textit{ReLU}. 
For the embedding model, we use \textit{Doubao-Embedding}, which is one of the Doubao models family from Bytedance.

\subsection{Evaluation of Tickets Escalating}
In this section, we systematically evaluate the performance of various ticket escalation methods. 
First, we conduct a comparative analysis of different approaches, including those based on Small Language Models (SLMs) and Large Language Models (LLMs). 
For LLM-based methods, we further evaluate prompt engineering and fine-tuning technologies.
Next, we perform ablation studies to investigate the effects of different data augmentation techniques specifically designed for methods that based on fine-tuning. 
Finally, we assess the performance of various base LLMs under two key paradigms: Chain-of-Thought (CoT) reasoning and In-Context Learning (ICL).

\subsubsection{Primary results}
Table \ref{tab:slm_llm_result} presents the primary comparative results between SLM-based and LLM-based methods, including approaches that based on prompt-engineering and fine-tuning. 
As shown in the table \ref{tab:slm_llm_result}, SLM-based methods, which embed messages into dense embedding vectors followed by traditional machine learning models, perform significantly worse than LLM-based methods, achieving a maximum F-score of only 80.1\%. 
This performance gap can be attributed to the following reasons.
On the on hand, the small language model is limited by the size of its parameters, which restricts its language comprehension capabilities.
On the other hand, non-end-to-end models may suffer from potential information loss.

In contrast, LLM-based methods exhibit significantly superior performance, primarily due to their outstanding content understanding and zero-shot task generalization capabilities. 
Using a Chain-of-Thought (CoT) prompt, these methods achieve approximately 82\% for both precision and recall. 
Moreover, employing a Reflection prompt, which enables the model to self-correct its reasoning and outputs, slightly improves precision to 82.8\%. 
We believe that one potential reason why the reflection technique does not significantly improve the experimental results is that the CoT prompting already makes the model with sufficient reasoning prior to reaching a conclusion.
During the reflection phase, the model does not acquire new information to help with generating more accurate or insightful outputs. 
Additionally, the In-Context Learning (ICL) prompt substantially increases recall to 89.2\%, albeit with a slight trade-off in precision. 
These results emphasize that the LLM-based methods process generalization capabilities and are effective tools for ticket escalation.

The application of supervised fine-tuning (SFT) to LLMs can further enhances performance. 
Specifically, fine-tuning significantly boosts the recall for the CoT prompt from 82.1\% to 91.2\%, while maintaining a similar precision of 81.8\%. 
This approach also achieves the highest F1-score (86.2\%) among all methods compared, demonstrating the effectiveness of fine-tuning in leveraging LLM capabilities for ticket escalation tasks. 
However, when SFT is combined with other prompt-based methods, such as Reflection and ICL, a slight decline in performance is observed. 
This is likely to be attributed to the SFT process has incorporated some samples from ICL or data with a similar distribution as training data, enabling it to learn the corresponding content during offline fine-tuning.

\subsubsection{Comparison of Data Augmentation methods}
While performing supervised fine-tuning (SFT) on LLMs, it is crucial to prepare the fine-tuning dataset to align with the prompt format intended for subsequent inference. 
However, directly using the raw messages and their corresponding labels for fine-tuning the LLM does not conform to the structural requirements of a CoT prompt.

As introduced in Section \ref{sec:finetune}, the dataset is collected online that leverage the outputs of the LLM, including both the reasoning process and the predicted class label. 
For samples where the prediction by LLM is correct, both the reasoning steps and label can be directly employed for SFT, as they naturally align with the CoT prompt structure. 
These datasets are labeled as \textbf{Correct}. However, for samples where the prediction is incorrect, the associated reasoning process cannot be directly used for SFT, as they reflect flawed reasoning that could degrade model performance if included in training.

To address this, we investigate different methods to preprocess datasets containing incorrect predictions:

\begin{itemize}
    \item \textbf{Wrong}: Retaining the original flawed reasoning  from the LLM and labeling it as wrong.
    \item \textbf{Revised}: Revising the reasoning process by prompting the LLM to generate the correct thought based on the ground truth label.
\end{itemize}

\noindent Additionally, we evaluate the performance of SFT using different data augmentation strategies, including:

\begin{itemize}
    \item \textbf{Raw}: Employing only the raw messages and labels without incorporating reasonings.
    \item \textbf{Correct}: Using only the samples with correct reasoning and corresponding labels for fine-tuning.
\end{itemize}

The results of these different data augmentation methods are presented in Table \ref{tab:diff_data_aug_result}, highlighting the impact of dataset augmentation on model performance. 
From the table, it is evident that SFT with thoughts significantly outperforms the one lacking such reasoning. 
In addition, SFT utilizing only the correct thoughts and labels yields the highest performance, surpassing that of methods utilizing incorrectly labeled thoughts.

\begin{table}[t]
\centering
\caption{Comparison of different data augmentation methods with SFT + CoT on Doubao-pro.}
\begin{tabular}{l|ccc}
\toprule
\textbf{Data Augmentation} & \textbf{Precision} & \textbf{Recall} & \textbf{F1-score} \\
\midrule
Raw (without thoughts) & 0.798 & 0.841 & 0.821\\
Correct & 0.813 & 0.851 & 0.831 \\
Correct and Wrong & 0.819 & 0.869 & 0.843 \\
Correct and Revised & 0.818 & \textbf{0.912} & \textbf{0.862} \\
Correct, Wrong and Revised & \textbf{0.831} & 0.856 & 0.844 \\
\bottomrule
\end{tabular}
\label{tab:diff_data_aug_result}
\end{table}

\subsubsection{Comparison of different base LLM models}
In this section, we evaluate different open-source LLM base models, especially \texttt{QWen2.5} and \texttt{Llama3.1} family.
We assess their performance using Chain-of-Thought (CoT) and In-Context Learning (ICL) prompting strategies.
The results of the evaluation are summarized in Table \ref{tab:diff_llm_result}.
From the results presented in the table, it is evident that ICL enhances the performance of CoT across all the models.
Additionally, models with a larger number of parameters (\texttt{QWen2.5-72B} and \texttt{Llama3.1-70B}) perform significantly better than those with fewer parameters (\texttt{QWen2.5-7B} and \texttt{Llama3.1-8B}). 
It is noteworthy that the \texttt{QWen2.5} family outputs higher precision but lower recall results, while \texttt{Llama3.1-70B} performs closely with that of the \texttt{Doubao-pro}.
A potential explanation for the suboptimal performance observed within the \texttt{QWen2.5} family may stem from the lack of prompt tuning during our experiments.
In conclusion, the findings suggest that open-source LLM base models can achieve competitive performance with those of the API-based model \texttt{Doubao-Pro}. This highlights the viability of open-source alternatives in the domain of large language models.

\begin{table}[t]
\centering
\caption{Comparison of different base LLM models.}
\begin{tabular}{c|l|ccc}
\toprule
\textbf{Methods} & \textbf{Base LLMs} & \textbf{Precision} & \textbf{Recall} & \textbf{F1-score} \\
\midrule
\multirow{4}{*}{\centering \shortstack{\textbf{CoT}}} 
 & Doubao-pro & 0.817 & 0.821 & 0.819 \\
 & Qwen2.5-7B & 0.916 & 0.352 & 0.508\\
 & Qwen2.5-72B & 0.855 & 0.664 & 0.747\\
 & Llama3.1-8B & 0.801 & 0.754 & 0.777 \\
 & Llama3.1-70B & 0.825 & 0.808 & 0.816\\
\midrule
\multirow{4}{*}{\centering \shortstack{\textbf{CoT + ICL}}} 
 & Doubao-pro & 0.810 & 0.892 & 0.849\\
 & Qwen2.5-7B & 0.836 & 0.491 & 0.618\\
 & Qwen2.5-72B & 0.837 & 0.781 & 0.808\\ 
 & Llama3.1-8B & 0.804 & 0.813 & 0.809\\
 & Llama3.1-70B & 0.809 & 0.885 & 0.845\\ 
\bottomrule
\end{tabular}
\label{tab:diff_llm_result}
\end{table}

\subsection{Evaluation of Escalation deduplication}

In order to reduce the duplicate ticket escalations caused by the same problem, we propose how to identify duplicate tickets in Section \ref{subsec:escalation_deduplication}. 
In the ticket escalation process, we employ representation vectors for each ticket to evaluate the similarity between different ticket issues. 
In the experiment, we conduct experiments to investigate the impact of the threshold parameter \(\theta\) on ticket deduplication. 
We then conduct an ablation study on the rewriting process, aimed at confirming its beneficial effects on the reduction of ticket escalations.

\subsubsection{Threshold parameter selection}
\label{subsubsec: deduplication_threshold}

In our labeled escalated tickets, we regard each escalated ticket as a separate category, and 14.7\% of the tickets are in linked states and considered as duplicated escalations.
To effectively assess the varying impacts of the similarity threshold \(\theta\) on ticket deduplication, we employ Precision, Recall and F1-score for evaluation under different threshold settings.

\begin{table}[t]
    \caption{Evaluation metrics of escalation deduplication under different threshold $\theta$.}
    \centering
    \begin{tabular}{c|ccc}
    \toprule
    Threshold (\(\theta\)) & \textbf{Precision} & \textbf{Recall} & \textbf{F1-score} \\
    \midrule
    0.86 & 0.945 & 0.806 & 0.865 \\
    0.87 & 0.934 & 0.826 & 0.870 \\
    \textbf{0.88} & 0.932 & 0.847 & \textbf{0.879} \\
    0.89 & 0.929 & 0.849 & 0.877 \\
    0.90 & 0.925 & 0.845 & 0.870 \\
    0.91 & 0.921 & 0.842 & 0.866 \\
    0.92 & 0.918 & 0.838 & 0.861 \\
    0.93 & 0.911 & 0.830 & 0.851 \\
    0.94 & 0.911 & 0.830 & 0.850 \\
    0.95 & 0.910 & 0.829 & 0.849 \\
    \bottomrule
    \end{tabular}
    \label{tab:deduplication_threthold}
\end{table}

Since the threshold $\theta$ represents the similarity between escalated tickets and their linked tickets, in this study, we report a reasonable range for the threshold between 0.86 and 0.95.
In Table \ref{tab:deduplication_threthold}, we evaluate the Precision, Recall and F1-score for escalation deduplication under different threshold parameter $\theta$.
We can find that F1-score initially increase as $\theta$ gradually raise, followed by a subsequent decline.
We attribute this phenomenon to two primary factors.
First, \name{} uses the embedding model as a zero-shot model, as illustrated in Figure \ref{fig:deduplication}.
When the threshold is set to a high value, the strict similarity constraints may result in similar issues being classified into disparate categories. 
This leads to an inflated count of tickets categorized as escalated in the evaluation relative to the ground truth, which does not show a monotonic trend with respect to the threshold $\theta$ during the evaluation. 
Second, it is important to acknowledge the inherent limitations of ticket deduplication based on ticket issues.
For tickets that exhibit the same symptoms but have different root causes, this method may lead to incorrect deduplication.
Table \ref{tab:deduplication_threthold} shows that the F1-score reaches its maximum of 0.879 when $\theta$ is set to 0.88.
We believe this threshold can help us achieve our goal of escalation deduplication, and it has been successfully implemented within our online production environment.

\subsubsection{Ablation study on ticket issue rewriting}

\begin{table}[t]
    \centering
    \begin{threeparttable}
    \caption{Ablation study on the ticket issues rewriting.}
    \begin{tabular}{l|ccc}
    \toprule
    & \textbf{Precision} & \textbf{Recall} & \textbf{F1-score} \\
    \midrule
    Escalated only \tnote{a} & 0.909  & 0.837 & 0.864 \\
    Escalated + rewriting \tnote{a}     & 0.932 & 0.847 & 0.879 (1.7\%$\uparrow$) \\
    \midrule
    Escalated only \tnote{b}     & 0.926 & 0.621 & 0.706\\
    Escalated + rewriting  \tnote{b}  & 1.000 & 0.642 & 0.749 (6.1\%$\uparrow$)\\
    \bottomrule
    \end{tabular}

    \begin{tablenotes}
        \footnotesize
        \item[a] All escalated tickets.
        \item[b] Escalated tickets that have more than one linked tickets.
    \end{tablenotes}
    \label{tab:deduplication_ablation}
    \end{threeparttable}
\end{table}

\name{} rewrites the issue of the tickets when it identifies duplicate escalations as Section \ref{subsec:escalation_deduplication} describes.
The rewritten ticket issue can comprehensively represent the phenomenon of similar issues, thereby avoiding the potential bias that specific tickets may fail to adequately encapsulate the essence of this type of issue.

We first use all the labeled dataset conduct ablation experiments to study the impact of rewriting.
In Table \ref{tab:deduplication_ablation}, the experimental results show that \name{} with rewriting achieves a 1.7\% improvement in F1-score compared to the setting without rewriting.
We note that \name{} performs escalation deduplication with an online manner.
It only rewrites the ticket issue when the escalated ticket has at least one linked ticket.
In other words, the advantages of rewriting may not be evident in scenarios involving only two similar tickets related to a specific issue.
To further refine our analysis, we refine the dataset by removing such cases, retaining only the escalated tickets that contain more than one linked ticket, and conduct experiments accordingly.
The experimental results show that the design of rewriting helps the F1-score of escalation deduplication improve from 0.706 to 0.749, representing a 6.1\% improvement attributable to the design of the rewriting mechanism.

\begin{figure}[t]
    \centering
    \includegraphics[width=\linewidth]{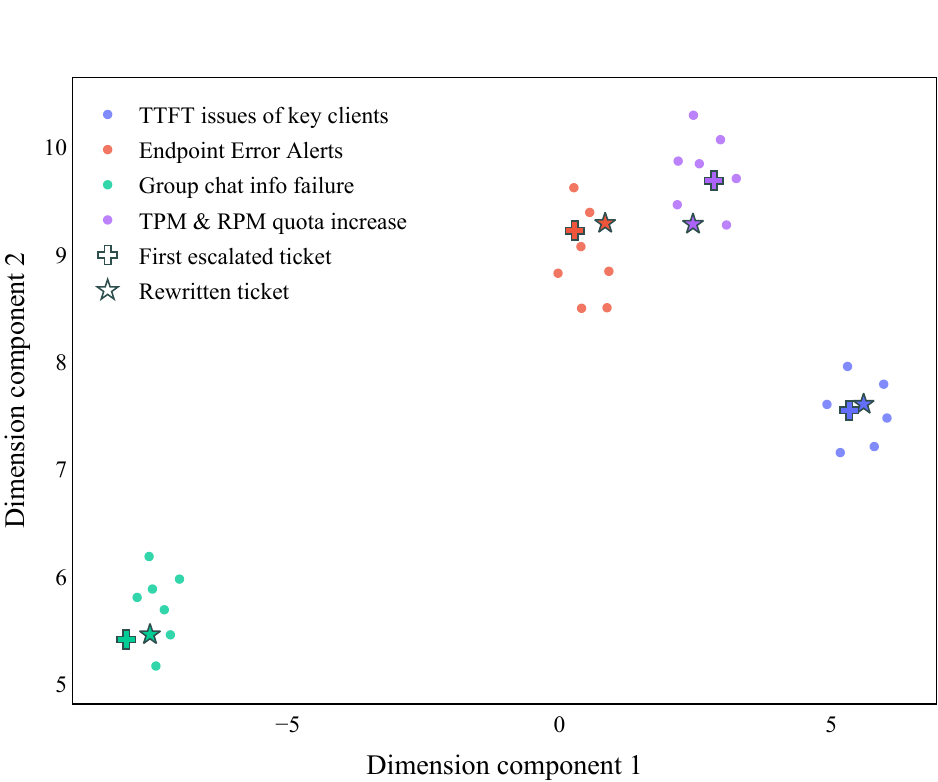}
    \caption{Samples of visualized embeddings for escalated tickets with dimensionality reduction.}
    \label{fig:deduplication_visualization}
\end{figure}

To visually illustrate the relationship between the rewritten ticket issues and their original tickets, we sample four groups of escalated tickets. 
The embeddings which are reduced in dimensionality using UMAP\cite{mcinnes_umap_2020} are displayed in Figure \ref{fig:deduplication_visualization}.
The visual representation reveals that the embeddings of the rewritten issues remain closely clustered within the categories of their original issues.
The above experiments indicate that the rewriting design in \name{} can correct the representation of similar issues, leading to more accurate deduplication of ticket escalations.

\section{Discussion and Learned Lessons}
\label{sec:discussion}

We conduct a comprehensive analysis of the customer tickets that experienced erroneous escalations by \name{}.
In this section, we present our observations and discuss the lessons learned.

\textbf{Limitations of Semantic-Based Ticket Escalations.}
In \name{}, the large language model assists in analyzing the content of tickets and determining whether an escalation is necessary. However, it still has several limitations.
First, we notice that personalized expressions from different customers can influence the performance of the LLM.
Exaggerating the impact of issues by customers may lead to inappropriate escalations, while a bland description of serious problems could result in \name{} overlooking necessary escalations.
Secondly, similar ticket descriptions may reflect different levels of severity depending on the specific cloud service products involved.
When users fail to clearly associate their issues with specific cloud services, it may result in erroneous escalations. 
For instance, a statement like "Multiple instances have encountered errors" could indicate a critical issue when referring to foundational infrastructure services, such as GPU instances.
However, if the ticket pertains to tasks within a data platform, such as offline Spark tasks, it often simply indicates that the customer has experienced multiple retry failures.
Therefore, when users submit tickets without clearly specifying the associated cloud service products, it can lead to misunderstandings by the LLM, resulting in incorrect escalations of the tickets.
This is the reason that we encourage the LLM to identify cloud service products from the tickets, as illustrated in Figure \ref{subfig:summary_prompt}.

\textbf{\name{} enhances human efficiency and reduces MTTR.}
Within our cloud platform, Volcano Engine, \name{} processes hundreds of tickets daily, with each ticket containing an average of over thirty dialogue records.
We estimate that reviewing all messages within each ticket and determining whether escalation is necessary takes approximately one minute for a human analyst.
By analyzing a sample of one hundred tickets, it is estimated that \name{} saves approximately ten person-days costs per day. 
Furthermore, this efficiency is expected to increase linearly with the growing number of tickets.
Since the launch of \name{}, we have collected data on the average resolution time for tickets escalated via \name{} compared to tickets escalated manually.  
The results indicate that \name{} reduces the mean time to repair (MTTR) by 39\%.
We attribute this improvement to the ability of \name{} to automatically analyze ticket content with higher accuracy compared to manual analysis.  
This prevents delays in addressing critical issues, thereby contributing to overall improved performance.

\textbf{Feedback and suggestions from support analysts.}
Since its launch, \name{} has gained significant recognition from support analysts.
They have noted that \name{} aids in the accurate and timely identification of issues within tickets, effectively preventing potential losses.  
Meanwhile, they have also provided some suggestions for improvement.
One notable suggestion pertains to the configurable ticket types.  
Although \name{} provides default ticket escalation types for all tickets, as shown in Figure \ref{fig:escalation_prompt}, this facilitates unified management and ensures the basic effectiveness of ticket escalation.
Some support analysts have expressed a desire to customize and subscribe to specific ticket types tailored to their needs.  
We believe that this flexibility could further enhance the ticket escalation process.  
However, the diversity of prompts might introduce instability in the performance of the large language model.  
Therefore, maintaining the accuracy of ticket escalations through backtesting after modifications to subscription ticket types represents a valuable direction for future research.

\section{Conclusion and Future Work}
\label{sec:conclusion} 
In this paper, we propose \name{}, a method for the escalation of customer tickets within the cloud platform Volcano Engine.  
\name{} leverages advanced large language models to accurately comprehend the content of incoming tickets and determine whether they belong to predefined ticket types for escalation.  
It also incorporates a deduplication process to minimize redundant escalations for similar tickets.  
We conduct a series of extensive experiments to evaluate the precision and recall of \name{} in ticket escalation.  
Furthermore, we introduce a category-guided fine-tuning methodology aimed at enhancing the overall performance of the model.

Since its launch, \name{} has significantly improved operational efficiency and received recognition from numerous support analysts.  
By analyzing its online performance, we identify some limitations of \name{}.  
In our future work, we plan to explore additional perspectives for ticket escalations and assignments, such as distributing tickets based on customer tiers or specific cloud service products.  
Furthermore, we aim to investigate configurable ticket escalation subscriptions to enhance flexibility and adaptability.

\bibliographystyle{ACM-Reference-Format}
\bibliography{ticket}
\end{document}